\documentclass[amsmath,twocolumn,pre]{revtex4}
\usepackage{graphicx}% Include figure files
\usepackage{dcolumn}% Align table columns on decimal point
\usepackage{subfigure}
\usepackage{latexsym,amssymb,amsmath}
\usepackage{bm}% bold math

\begin{document}

%\preprint{APS/123-QED}

\title{Jamming in granular hopper flow}

\author{James W. Landry}
 \email{james.landry@baesystems.com}
 \affiliation{BAE Systems Advanced Information Technologies, Burlington,
   Massachusetts 01803}
\author{Gary S. Grest}
 \affiliation{Sandia National Laboratories, Albuquerque, New Mexico 87185}

\date{June 25, 2003}% It is always \today, today,

\begin{abstract}
  Large-scale three dimensional molecular dynamics simulations of hopper flow
  are presented.  The flow rate of the system is controlled by the width of
  the aperture at the bottom.  As the steady-state flow rate is reduced, the
  force distribution $P(f)$ changes only slightly, while there is a large
  change in the impulse distribution $P(i)$.  In both cases, the distributions
  show an increase in small forces or impulses as the systems approach
  jamming, the opposite of that seen in previous Lennard-Jones simulations.
  This occurs dynamically as well for a hopper that transitions from a flowing
  to a jammed state over time.  The final jammed $P(f)$ is quite distinct from
  a poured packing $P(f)$ in the same geometry.  The change in $P(i)$ is a
  much stronger indicator of the approach to jamming.  The formation of a peak
  or plateau in $P(f)$ at the average force is not a general feature of the
  approach to jamming.
\end{abstract}

%\pacs{75.50.Lk, 75.10.Nr}% PACS, the Physics and Astronomy
%\keywords{spin glass; ground states; two-dimensional; polynomial-time
%algorithm}
\maketitle

\section{Introduction}

The vertical pressure of grains in a silo becomes independent of depth for
sufficiently tall silos~\cite{Janssen1895,JaegerOct1996,VanelFeb2000}.  The
weight of sand in the column is ultimately supported by the lateral containing
walls.  Experiments~\cite{HowellJun1999,GengJuly2001,BlairApr2001,LiuJul1995},
theory~\cite{CoppersmithMay1996}, and
simulations~\cite{MakseMay2000,SilbertDec2002,LandryApr2003} all show that
these stresses are transmitted in an inhomogeneous manner.  One measure of the
inhomogeneity in a granular pile is expressed through the distribution of
normal forces, $P(f)$, where $f = F/\langle F \rangle$ is the normalized
force, $F$ is the normal force, and $\langle F \rangle$ is the average force.
$P(f)$ of static granular piles exhibits an exponential tail at large forces
and a plateau or turnover at small
forces~\cite{BlairApr2001,LiuJul1995,EriksonOct2002,LoevollNov1999,BrujicJun2002,SilbertDec2002,SnoeijerJul2004,MetzgerNov2004}.
A fundamental question is whether these inhomogeneities are present in flowing
granular material.  In addition, if these inhomogeneities are distinct in the
flowing state, then it may be possible to predict the approach of the quiescent
or ``jammed'' state by analyzing the change in inhomogeneities as one moves from
one to the other.  In the context of a general conception of
jamming~\cite{Liu1998}, it has been conjectured~\cite{OhernJan2001} that the
$P(f)$ of a flowing system exhibits a \textit{decrease} at small forces ($f <
1$) as the system moves from a flowing to a jammed state.
 
An ideal system to probe this transition is the hopper~\cite{ZhuMar2005}.  Its
jammed state (a silo) has been extensively studied.  For a granular packing of
sufficient height, the pressure in the depth of the packing is independent of
the height of the granular matter above it~\cite{Janssen1895}.  A granular
silo is also constrained and well-controlled --- there is no free surface
aside from the top.  A silo transitions to a hopper when an aperture at the
base is opened.  Numerous experiments and simulations have examined the
transition to jamming with the goal of predicting its
onset~\cite{ZuriguelSep2003,ToJul2002,ToJan2001}.  Here we focus specifically
on the distribution of forces and impulses in the approach to jamming.

In experiments the forces between particles are difficult to measure
directly~\cite{EaswarNote2}, so Longhi {\it et al}~\cite{LonghiJul2002}
measured the distribution of impulses $P(i)$ (where the normalized impulse is
$i = I/\langle I \rangle$, $I$ is the impulse, and $\langle I \rangle$ is the
average impulse) between particles and the wall and related this distribution
to the macroscopic behavior of the system as the system approached jamming.
This quasi-$2D$ experiment found that the impulse has an exponential tail at
all flow rates and that at small flow rates, the distribution of $P(i)$
increases with decreasing flow rate.  The distribution of times between
collisions tends to a power law $P(\tau_c) \approx \tau_c^{-3/2}$, that is,
the mean time interval tends to diverge just as in the glass
transition~\cite{LonghiJul2002}.  Ferguson {\it et al} found using $2D$
event-driven frictionless simulations~\cite{FergusonFeb2004} that $P(i)$
exhibits an \textit{increase} in small impulses ($i < 1$) and argue that this
arises from the contribution of ``frequently-colliding'' particles that are
spatially correlated into $1D$ collapse strings.

Here we present fully $3D$ molecular dynamics simulations of hopper
flow.  We measure the distributions of normal forces $P(f)$ and impulses
$P(i)$ for the system as the system approaches jamming and relate our
results to previous experiments and $2D$ simulations.  We find that
$P(f)$ is not a strong indicator of jamming, and that it exhibits an
\textit{increase} at small forces as the system approaches jamming,
exactly the opposite of the predicted behavior from earlier
simulations~\cite{OhernJan2001}.  $P(f)$ of a jammed system is markedly
different from a poured packing in a similar geometry, showing that
$P(f)$ can depend strongly on the history of the system.  We also find
that $P(i)$ exhibits an \textit{increase} at small impulses, in
agreement with previous experiments~\cite{LonghiJul2002} and
simulations~\cite{FergusonFeb2004}.

\section{Simulation Method}

\begin{figure*}
\includegraphics[height=4.0in]{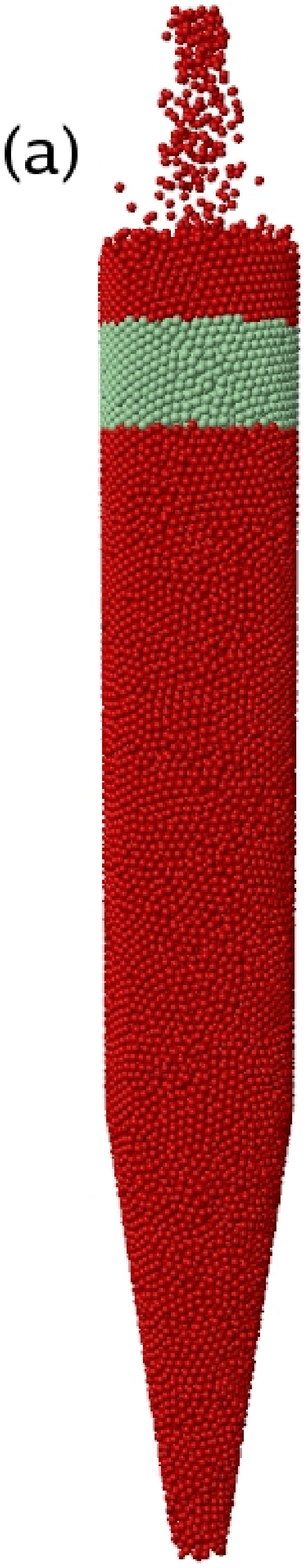}
\includegraphics[height=4.0in]{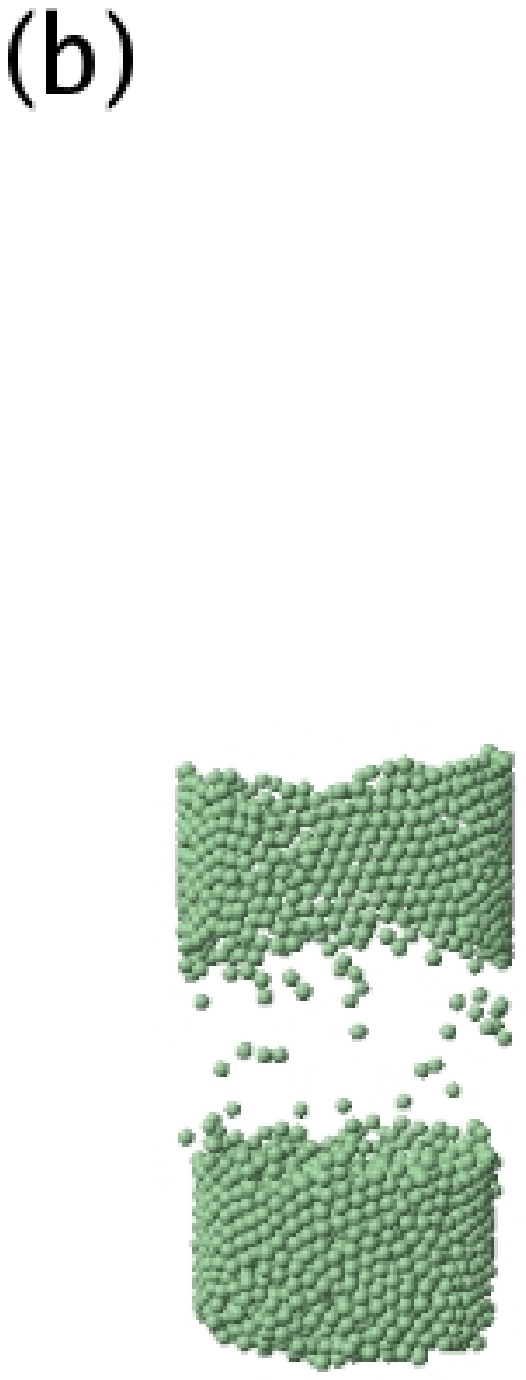}
\includegraphics[height=4.0in]{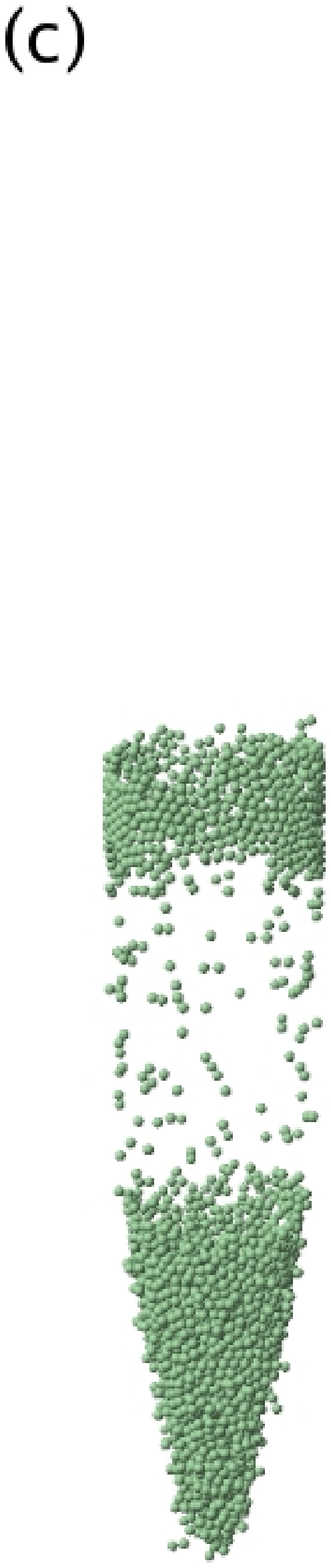}
\includegraphics[height=4.0in]{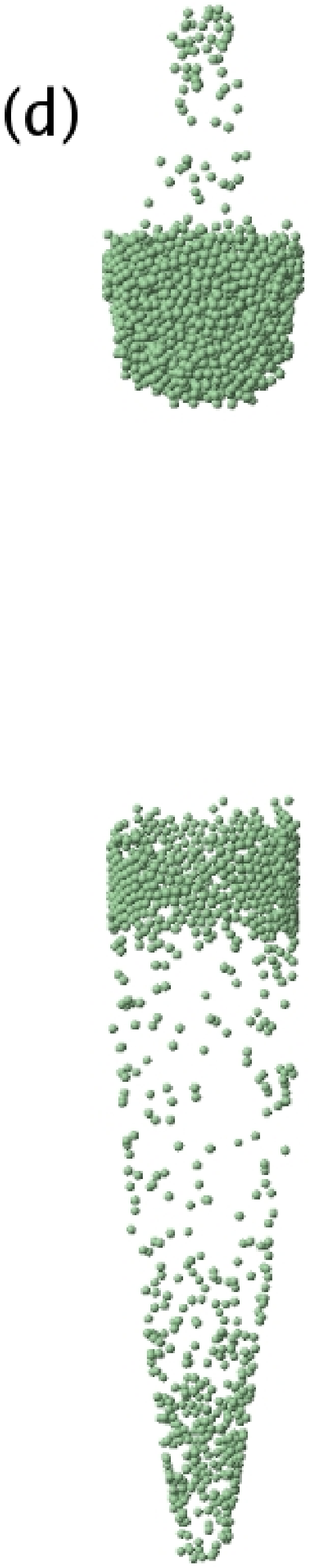}
\caption{\label{fig:diagram}(color online) (a) Geometry of the system.
  The total height of the system is $160d$, with the freefall region
  roughly $20d$ in height, the cylindrical region $90d$, and the funnel
  region $50d$.  The bottom aperture radius $r_f$ is varied to control
  the flow.  In this case, $r_f = 3d$.  All particles with $z$ positions
  $120d < z < 130d$ are colored gray (green).  (b) Only the tracer
  particles after time $t = 300 \tau$.  (c) Tracer particles only at $t =
  580 \tau$.  (d) Tracer particles only at $t = 680 \tau$.}
\end{figure*}

Our geometry was inspired by that used in the quasi-$2D$ experiments: a
conical hopper~\cite{Nedderman1992}.  The system is periodic in the $z$
direction: particles that fall out of the bottom of the system reappear
at the top and refill the container, creating a steady-state flowing
system.  The system is shown in Figure~\ref{fig:diagram}(a).  The top
of the hopper is a cylindrical container with radius $R = 10d$, where
$d$ is the particle diameter.  At the bottom of this cylindrical region,
a cone is joined to the cylinder, with radius varying smoothly from $R$
to $r_f$ over a distance $z_f = 50d$.  The angle $\theta$ between
vertical and the funnel is $\theta = \tan^{-1} \left(\frac{R -
r_f}{z_f}\right)$.  This system tends to jam for $r_f < 1.5d$.  We
deliberately make the cylindrical region very deep to assure that in the
static case the pressure is independent of depth.  Those particles that
leave the hopper opening and then rain down on the top of the pile
quickly lose their kinetic energy and do not affect the flow through the
opening.  The system is prepared by pouring particles into the top of
the container with aperture radius $r_f = d$ and a plate at the bottom to
prevent the exit of particles.  After the packing is formed, the plate is
removed and the aperture widened to the desired radius.  All
measurements are taken after the system has cycled particles through at
least once and is in a steady state unless specifically noted otherwise.
We focus our attention on the funnel region, and unless otherwise
stated, all measurements are only of particles in that region.

We use a molecular dynamics code with a Hertzian force law that is
described in detail elsewhere~\cite{LandryApr2003,SilbertOct2001}.  All
results will be presented in dimensionless units based on the mass of a
particle $m$, the diameter of a particle $d$, and the force of gravity
$g$.  In this case, the simulations used $N = 40000$ monodisperse
particles of diameter $d$.  The unit of time is $\tau = \sqrt{d/g}$, the
time it takes a particle to fall its radius from rest under gravity.
Most of the simulations were carried out with a spring constant $k_n = 2
\times 10^5 mg/d$, for which the simulation time step $\delta t =
10^{-4} \tau$. The particle-particle friction and particle-wall friction
are the same: $\mu = \mu_w = 0.5$.

% remember to reference paper where they showed stuff was eaten away

\begin{figure}
\includegraphics[width=3in,clip]{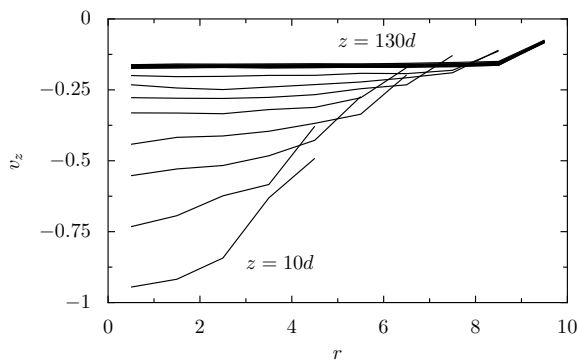}
\caption{\label{fig:flowprof}Velocity profiles for $v_z$ at a number of
  heights $z$ in a flowing hopper with $r_f = 3d$.  The profile with the
  largest magnitude was taken at $z = 10d$, and at every $5d$ afterward,
  up to a maximum of $130d$.  $|v_z|$ decreases as the funnel widens
  with increasing height, until one reaches the cylindrical region,
  where it is roughly constant.}
\end{figure}

The flow profile is dependent on the geometry of the system, as shown in
Fig.~\ref{fig:flowprof}.  The flow is essentially plug-like and constant
in the cylindrical region of the hopper for particles more than one
particle away from the wall.  A one-particle thick layer near the wall
flows more slowly due to friction with the wall.  This plug breaks up in
the funnel region, with particles in the center of the flow flowing more
quickly and accelerating as they approach the hopper opening.

Experimental measurements have shown that hoppers produce a mass flow
rate dependence of $M = (r_f - wd)^{5/2}$, where $w$ is a constant
related to the shape of the particles~\cite{NeddermanNov1982a}. We see a
similar dependence in our simulations, with $w = 1.2$, consistent with
the experimental measurement of $w \approx 1.5$~\cite{NeddermanNov1982a}.

\section{Forces and Impulses}

\begin{figure}
\includegraphics[width=2.5in,angle=270,clip]{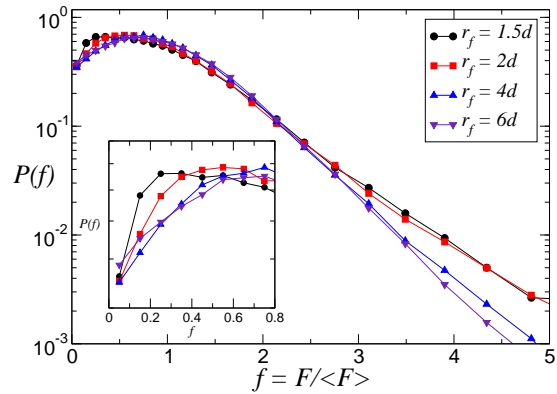}
\caption{\label{fig:pof}Distribution of normal forces $P(f)$ for flowing
  hopper systems with various apertures $r_f$.  $P(f)$ increases
  slightly at small forces as the system approaches jamming.  Inset:
  closeup of $P(f)$ at small forces.  The lines are guides to the eye.}
\end{figure}

The distribution of particle-particle normal forces $P(f)$ for the
funnel region is shown in Fig.~\ref{fig:pof} for a variety of aperture
radii $r_f$.  $P(f)$ is exponential at large forces, just as observed in
static
piles~\cite{BlairApr2001,LiuJul1995,EriksonOct2002,LoevollNov1999,BrujicJun2002}.
All the flowing states of the system have similar distributions.  In
Fig.~\ref{fig:pof} (inset) we show a closeup of $P(f)$ for a range of
flowing states.  There is a slight change in $P(f)$ as the system
approaches jamming --- it displays a slight \textit{increase} at small
forces.  The peak of the distribution moves to smaller forces as the
system approaches jamming.  This behavior is the opposite of the
behavior seen in $2D$ Lennard-Jones simulations~\cite{OhernJan2001}.
This change in $P(f)$ is robust and suggests that the behavior of $P(f)$
in a jammed system is dependent on the specific physics of the
interactions between particles.

\begin{figure}
\includegraphics[width=2.5in,angle=270,clip]{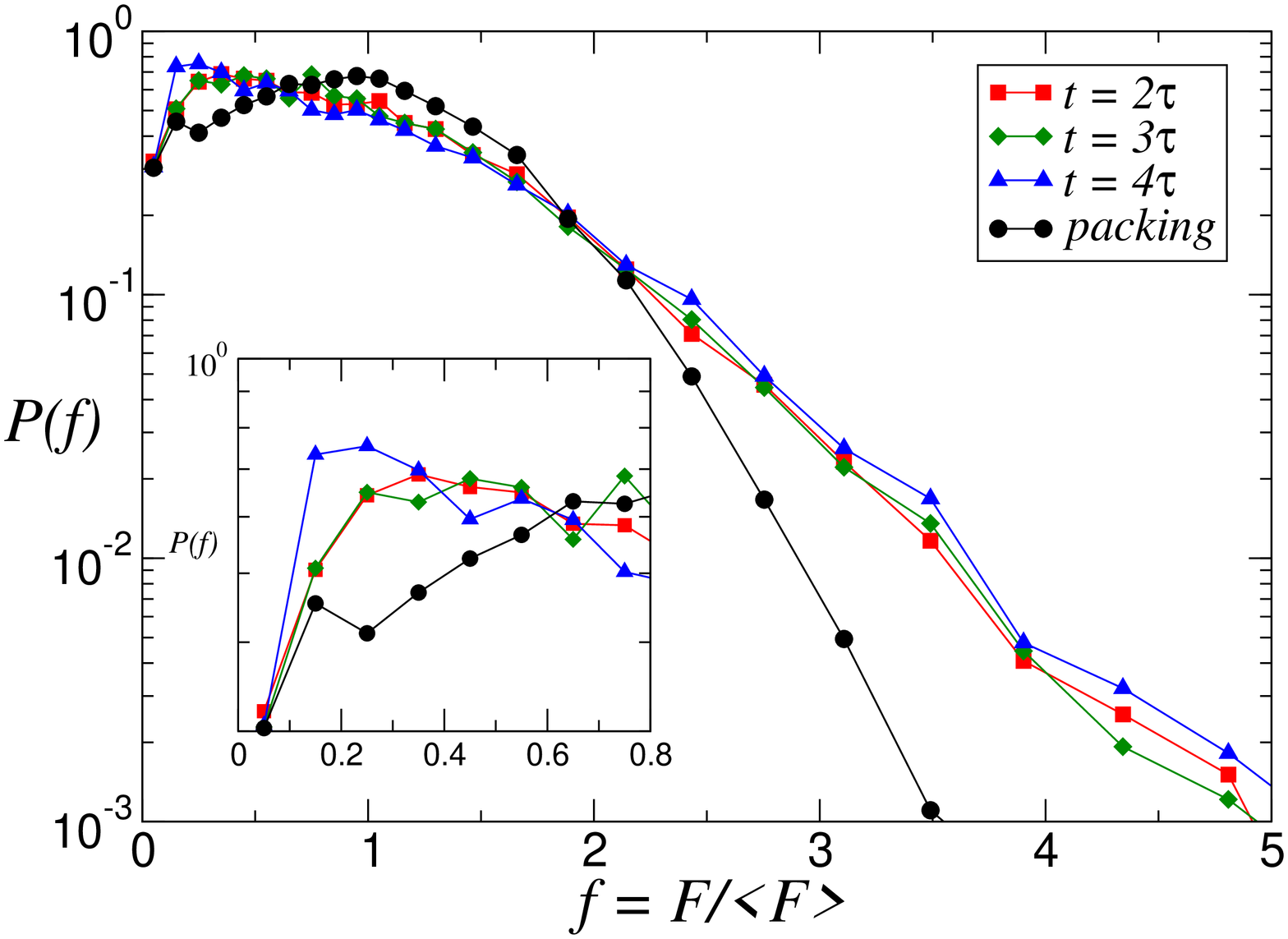}
\caption{\label{fig:pof-time}Distribution of normal forces $P(f)$ for
  flowing hopper systems with aperture radius $r_f = 1.4d$ as a function
  of time.  The system flow freely until $t \sim 3.6 \tau$, when a jam
  develops and shifts the system towards an increase in $P(f)$.  The
  jammed $P(f)$ is distinct from that of a poured state $P(f)$, here
  shown for a system that never flows after the aperture is opened at
  $r_f=d$.  Inset: closeup of $P(f)$ at small forces.}
\end{figure}

We also analyzed the dynamical nature of $P(f)$ for $r_f = 1.4d$.  In
this case, the system flows freely for a short period and then jams.  In
Fig.~\ref{fig:pof-time} $P(f)$ is shown for this system as a function of
time.  As the jam forms, the distribution abruptly increases at small
forces~\cite{FlowApt1}.  This behavior is compared
to that of a poured system with $r_f = d$ that never flows after the
aperture is opened.  In that case, $P(f)$ is lower at small forces than
the flowing state.  The formation of a jam is history-dependent, and the
static packing that results from a jam is fundamentally distinct from a
poured packing.  $P(f)$ in a jammed packing is strongly dependent on the
dynamic processes used to create the jam and is not equivalent to the
$P(f)$ of a granular packing created with an alternate method.  There is
no general $P(f)$ that one should expect for all jammed packings.

This hysteretic sudden change in $P(f)$ has also been seen in reverse in
shearing experiments in cylinders.  In that case, $P(f)$ is essentially
unchanged until yield stress is attained, at which point $P(f)$ changes
discontinuously to a new form~\cite{CorwinNote1}.

\begin{figure}
\includegraphics[width=2.5in,angle=270,clip]{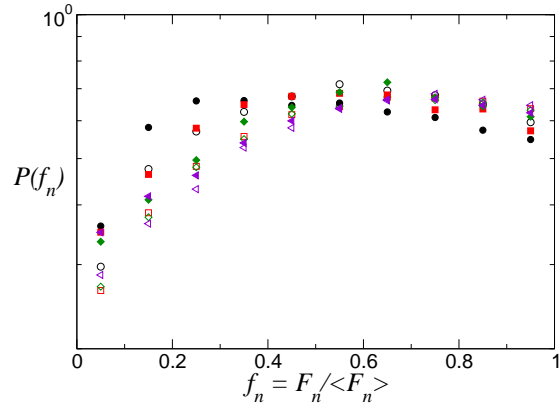}
\caption{\label{fig:pof-spring}Distribution of normal forces $P(f)$ for
  flowing hopper systems with various aperture radius $r_f$.  As in the
  previous plots, the symbols represent aperture radii: $\bigcirc$
  $r_f=1.5d$, $\Box$ $r_f=2d$, $\Diamond$ $r_f=3d$, and $\triangleleft$
  $r_f=5d$.  Filled-in symbols represent stiff springs of $k_n = 2
  \times 10^5 mg/d$ and open symbols represent loose springs of $k_n = 2
  \times 10^3 mg/d$.  Soft springs show a smaller effect in the approach
  to jamming.}
\end{figure}

We can also probe the dependence of $P(f)$ on the particle interactions.
We carried out additional simulations with a much softer normal spring
constant, $k_n = 2 \times 10^3 mg/d$.  Experiments have shown that very
soft particle interactions can have a strong effect on
$P(f)$~\cite{EriksonOct2002}.  We compare the behavior of $P(f)$ for the
two systems in Figure~\ref{fig:pof-spring}.  The trend of increasing the
$P(f)$ at small forces is the same in both systems, but the increase is
more pronounced and more localized at small forces for the stiffer
springs.  $P(f)$ is thus sensitive to the compressibility of
the particles and softness of the particle interactions.

These analyses provides a clue to the discrepancy between our
simulations and those of O'Hern {\it et al}~\cite{OhernJan2001} and
others.  Two possible effects separate our simulations and theirs.  The
first and more important effect arises from the ahistorical nature of
the O'Hern simulations, in which the packings are produced by conjugate
gradient methods, which have no information on the dynamics of the
transition from flowing state to jammed state.  As demonstrated above,
in granular materials the history of the system cannot be ignored, and
the $P(f)$ of a jammed state is quite distinct from that of a poured
state.  The $P(f)$ obtained for the repulsive interactions by O'Hern
{\it et al} more closely resemble those for static granular packings and
are in fact produced by the quasi-static method of conjugate gradient.
In our simulations a poured packing does exhibit a deficit at small
forces in relation to flowing systems, but all ``jammed'' systems clearly do
not exhibit the same characteristics.  A poured packing is not jammed in the
same sense as a plugged hopper flow packing.

The second effect is that some simulations use an extremely soft
potential that allows significant overlap.  As shown above, softer
potentials tend to diminish the increase at small forces observed in our
simulations.  For sufficiently soft potentials, $P(f)$ could have a
deficit at small forces.  Thus, we do not expect our findings to agree
with simulations undertaken with extremely soft potentials.

\begin{figure}
\includegraphics[width=2.5in,clip,angle=270]{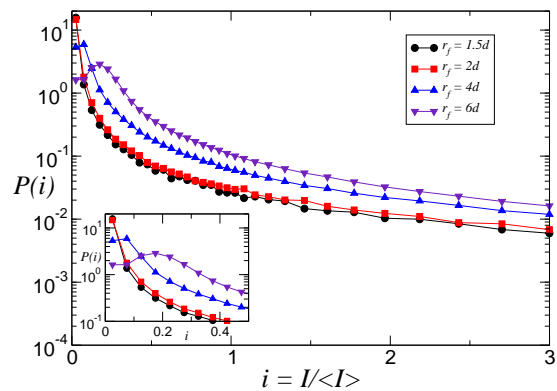}
\caption{\label{fig:poi}Probability distribution of impulses $P(i_n)$
  for the funnel region.  The system exhibits a strong \textit{increase}
  in small impulses as the system approaches jamming.  Inset: closeup of
  $P(i_n)$ at small impulses.}
\end{figure}

To make contact with experiment~\cite{LonghiJul2002} and
simulation~\cite{FergusonFeb2004,DennistonMar1999}, we also analyze the
impulse distribution.  Figure~\ref{fig:poi} shows the distribution of
particle-particle normal impulses $P(i)$ for contacts that terminate in
the hopper region~\cite{ImpulseStatic1}.  The tails of the system are no
longer exponential; they decline faster than
exponential~\cite{ForceTails1}~\cite{EaswarNote1}.  There is also a
strong signal as the system approaches the jammed state: $P(i)$
\textit{increases} at small impulses ($i = I/\langle I \rangle < 1$).
This increase at small impulses is also seen in the quasi-$2D$ hopper
experiments~\cite{EaswarNote1,LonghiJul2002}.

Our measurements were begun after the particles had already circulated
through the container at least once, and impulses were calculated for a
duration of $25 \tau$, which is greater in duration than essentially all
contacts.  Impulses were calculated for all those contacts that ended in
the funnel region.  These results are somewhat consistent with the
change of $P(i)$ from $2D$ event-driven
simulations~\cite{FergusonFeb2004}.  In the $2D$ simulation, the change
in $P(i)$ at small impulses was controlled by ``rapidly colliding
particles'', which formed 1D linear ``collapse strings''.  We denote
``rapidly colliding particles'' in our case as those with the largest
number of collisions over a set time window, but we observe no obvious
spatial correlation of these particles.  

\section{Conclusions}

In summary, our simulations show that for a granular hopper system with
history-dependent interactions and friction, $P(f)$ exhibits an increase in
small forces $(f < 1)$ as the system approaches jamming.  This increase is
robust and hysteretic; the $P(f)$ of a jammed packing is distinct from the
$P(f)$ of a poured packing.  The behavior of $P(f)$ is somewhat dependent on
the interaction: softer particle interactions tend to diminish this effect.
The history dependence of granular materials distinguishes our simulations
from many $2D$ simulations{\it et al}~\cite{OhernJan2001}, which have no
memory of dynamics, and thus predict a general behavior of $P(f)$ that is
exactly the opposite of that observed in our simulations.  The trend is the
same for the distribution of impulses $P(i)$, which see an increase in small
impulses $(i < 1)$ as the system approaches jamming.  This trend is also
observed in experiments and quasi-$2D$ simulations.  The behavior of $P(f)$
does not appear to be a general feature of jamming, but instead depends on the
particle interactions and the hysteretic quality of the system.

We acknowledge helpful discussions with E. Ben-Naim, B. Chakraborty,
S.N. Coppersmith, N. Easwar, A. Ferguson, S. Majumdar, N. Menon, S.  Nagel,
C. Olson Reichhardt, C. Reichhardt, L. Sadasiv, L. Silbert, and S. Tewari.  We
thank J. Lechman for a critical reading of the manuscript.  This work was
supported by the Division of Materials Science and Engineering, Basic Energy
Sciences, Office of Science, U.S.  Department of Energy.  This collaboration
was performed under the auspices of the DOE Center of Excellence for the
Synthesis and Processing of Advanced Materials.  Sandia is a multiprogram
laboratory operated by Sandia Corporation, a Lockheed Martin Company, for the
United States Department of Energy's National Nuclear Security Administration
under contract DE-AC04-94AL85000.

\bibliography{grain}

\begin{thebibliography}{34}
\bibitem[{Eas({\natexlab{a}})}]{EaswarNote2}
\bibinfo{note}{In some cases, forces can be obtained through careful use of the
  impulse and timing data. Work is underway on this problem.}
\bibitem[{Flo()}]{FlowApt1}
\bibinfo{note}{Similar behavior was observed for other flowing apertures
  slightly less than $r_f = 1.5d$.}
\bibitem[{Imp()}]{ImpulseStatic1}
\bibinfo{note}{The distribution of impulses for an essentially static system is
  difficult to interpret.}
\bibitem[{For()}]{ForceTails1}
\bibinfo{note}{There is some evidence to suggest that the $P(f)$ distribution
  for flowing $3D$ experimental systems is also slightly non-exponential}.
\end{thebibliography}

\end{document}